\documentclass[pre, amssymb, amsmath, nofootinbib, 10pt, twocolumn]{revtex4}
	
	\usepackage{graphicx}
	\usepackage{color}
	\usepackage{verbatim}
	\usepackage{subfigure}
	\usepackage{amssymb}

\begin{document}
	
	\title{Nonequilibrium Heat Capacity}
	\author{Dibyendu Mandal$^{1,2}$}
	\affiliation{
	$^1$Department of Physics, University of Maryland, College Park, Maryland 20742, U.S.A. \\
	$^2$Department of Atmospheric and Oceanic Sciences, University of Colorado, Boulder, Colorado 80309, U.S.A. } 
	
	\begin{abstract}
	
	Development of steady state thermodynamics and statistical mechanics depends crucially on our ability to develop notions of equilibrium thermodynamics for nonequilibrium steady states (NESS). 
	The present paper considers the development of heat capacity. 
	A modified definition is proposed which continues to maintain the same relation to steady state Shannon entropy as in equilibrium, thus providing a thermodynamically consistent treatment of NESS heat capacity.
		
	\end{abstract}

	\maketitle

	Classical thermodynamics and statistical mechanics have been formulated for equilibrium states and transitions among them~\cite{Callen1985}. 
	They do not apply to nonequilibrium steady states (NESS), characterized by positive entropy production rate. 
	 But systems in NESS are ubiquitous in nature: from enzymes and molecular motors~\cite{Hill1977, Wang1998} and oscillating chemical systems~\cite{Qian2002} to virtually any system involving transport processes happening at a finite rate. 
	Therefore, to develop a consistent thermodynamic formalism for transitions between such states has been an ongoing program of intense research for more than half a century~\cite{deGroot1962, Glansdorff1971, Keizer1987, Jou1993, Oono1998, Sekimoto1998, Hatano2001, Ruelle2003, Gallavotti2004, Seifert2005, Komatsu2008, Esposito2010, VandenBroeck2010, Ge2010, Bertini2013}.
	Interest in the field has intensified further with the recent development of steady state fluctuation theorems~\cite{Evans1993, Evans1994, Gallavotti1995, Seifert2012}.
	However, the program is far from over as we are still in the process of developing the counterparts of equilibrium theory for NESS.
	In this paper we are concerned with the development of heat capacity.

	Heat capacity played a major role in the formulation of equilibrium thermodynamics and statistical mechanics. 
	It helped repeal the caloric theory of heat~\cite{Planck1990} and was one of the foremost quantities to be investigated in statistical mechanics~\cite{Huang1987}. 
	It is therefore interesting to see if this concept can be generalized to NESS. 
	Because of the constant entropy production rate (equivalent to a constant heat dissipation rate in the environment), however, heat capacity according to the regular definition, Eq.~\ref{eq:C}, is infinite for {\it any} NESS.  
	To circumvent this problem Boksenbojm {\it et al.}~\cite{Boksenbojm2011} utilized a heat normalization scheme proposed by Oono and Paniconi~\cite{Oono1998}. 
	In this scheme, one first considers the minimum amount of heat dissipation necessary to maintain the system in NESS, called the {\it housekeeping heat}, and then subtracts it from the total heat to obtain a normalized, {\it excess heat}. 
	By using this excess heat in place of the total heat Boksenbojm {\it et al.}~\cite{Boksenbojm2011} obtained a finite NESS heat capacity, Eq.~\ref{eq:CexOP}. 
	Unlike equilibrium, however, the heat capacity in this approach can not be written as the temperature derivative of a generalized thermodynamic potential~\cite{Boksenbojm2011}.
	Furthermore, the expression is model-dependent, involving its kinetic parameters explicitly, and thus, thermodynamically less attractive. 
	The purpose of the present paper is to give an alternate definition of NESS heat capacity that leads to a finite, model-independent expression, Eq.~\ref{eq:main}.

	Heat normalization scheme for NESS is not unique~\cite{Maes2013,Sasa2013arXiv}. 
	Hatano and Sasa~\cite{Hatano2001} proposed a normalization scheme which, unlike Oono and Paniconi's scheme~\cite{Oono1998, Komatsu2008, Bertini2013}, applied not only to nonequilibrium steady states (NESS) but also to the transient states. 
	The distinction between the two schemes seems to be an under-appreciated fact in the existing literature, leading to two different approaches to NESS thermodynamics under the same vocabulary~\cite{Maes2013}.
	Building on Hatano and Sasa's approach the present paper shows that an alternate definition of finite heat capacity is possible; furthermore, this new heat capacity is proportional to the temperature derivative of the steady state Shannon entropy (see Eq.~\ref{eq:main}) just as in the case of equilibrium states. 
	There are several proposals for nonequilibrium entropy~\cite{Maes2012}; however, for the general Markovian models considered below, Shannon entropy is the only candidate to satisfy all the following reasonable requirements~\cite{Mandal2013}: (1) it is finite for system with finite number of configurations; (2) it is a state function; and (3) it reduces to the equilibrium entropy if the system satisfies detailed balance~\cite{vanKampen2007}.  
	The present paper, therefore, gives a thermodynamically consistent treatment of NESS heat capacity.
	(At the end of the paper, developing on the following presentation, we provide further arguments to substantiate this claim.)
	In the following, the derivation of Eq.~\ref{eq:main} is first presented for a general Markovian jump process. 
	Then a specific diffusive process -- that of driven one dimensional diffusion with periodic boundary condition -- is considered as it has been the paradigmatic case of diffusive NESS systems~\cite{Seifert2012}.
	Generalization to higher dimensions is straightforward.

	{\bf Markovian jump processes. -- } 
	Consider a system making random, Markovian jumps among a finite number of physical configurations due to thermal fluctuations from a reservoir at absolute temperature $T$. 
	We can characterize these jumps by a set of nonnegative numbers $\{R_{ij}\}$ where $R_{ij}$ denotes the conditional rate of transition to configuration $i$ from configuration $j$. 
	These rates can be expressed in terms of the thermodynamic quantities of the system and the reservoir. 
	Let $E_i$ be the (free) energy of configuration $i$ and $B_{ij} = B_{ji}$ the (free) energy barrier between $i$ and $j$. 
	Whenever the system makes a transition from $j$ to $i$ an amount of heat $(E_j - E_i)$ is released in the reservoir leading to an increase in reservoir entropy by $(E_j - E_i) / T$.  
	In presence of nonconservative forces that drive the system out of equilibrium, additional amounts of entropy $\Delta S^{r, ext}_{ij} = - \Delta S^{r, ext}_{ji}$ are produced corresponding to the work done by these forces.  
	Transition rates $R_{ij}$ can then be written as~\cite{Seifert2012, Kolomeisky2007}
	\begin{equation}
	\label{eq:LDB}
	R_{ij} = \nu(T) \, \exp{\left[\frac{1}{k_B T}\left(- B_{ij} + E_j + \eta_{ij} T \Delta S_{ij}^{r, ext} \right) \right]},
	\end{equation}
where $\nu(T)$ is a frequency factor common to all the transitions, the numbers $\{\eta_{ij}\}$ satisfy the conditions $\eta_{ij} + \eta_{ji} = 1$, and $k_B$ is the Boltzmann constant. 
	In the following we shall omit the factors of $k_B$ for simplicity of expressions.

	Let $p_i(t)$ denote the occupation probability of any configuration $i$ at time $t$. 
	Because of the random transitions, $p_i(t)$'s evolve according to the master equation~\cite{vanKampen2007}
	\begin{equation}
	\label{eq:master}
	\frac{\mathrm{d}}{\mathrm{d}t} p_i (t) = \sum_{j \neq i} \Big[ R_{ij} p_i(t) - R_{ji}p_j(t) \Big].
	\end{equation}
	This can also be written as a continuity equation. 
	First, one introduces the instantaneous probability fluxes 
	\begin{equation}
	\label{eq:Flux}
	J_{ij}(t) = R_{ij} p_j(t) - R_{ji} p_i(t),
	\end{equation}
from any configuration $j$ to another configuration $i$.
	Then, the master equation~\ref{eq:master} is rewritten as
	 \begin{equation}
	\label{eq:MasterCurrent}
	\frac{\mathrm{d}}{\mathrm{d}t} p_i (t) = \sum_{j \neq i} J_{ij}(t), 
	\end{equation}
implying that the rate of change of probability $p_i(t)$ is simply the net probability flux into $i$. 
	When the rates are held fixed, the system relaxes to a steady state, $\{p_i^S\}$, which is unique if the system is both reversible and ergodic~\cite{vanKampen2007}. 
	(We assume this to be the case.) 
	Nonequilibrium steady states (NESS) differ from equilibrium steady states by having non-zero fluxes, $J_{ij}^S = R_{ij} p_j^S - R_{ji}p_i^S \neq 0$~\cite{Zia2007}.  
	These nonzero fluxes are the reason behind the positive heat dissipation rate in NESS~\cite{Ge2009}.

	Consider now initiating the system in its steady state at a temperature $(T - \delta T)$ and then bringing it instantaneously in contact with a reservoir at temperature $T$. 
	The system starts in the state $\{p_i(t = 0) = p_i^S(T - \delta T)\}$ and then relaxes to the state $\{p_i(\infty) = p_i^S(T)\}$ evolving according to Eq.~\ref{eq:master}. 
	If $\langle Q^r \rangle$ denotes the average dissipated heat in the reservoir during the relaxation process, the heat capacity in the ``equilibrium" scenario (i.e., all $J_{ij}^{S}$ are zero) is given by the relation 
	\begin{equation}
	\label{eq:C}
	C_{eq}(T) = -  \lim_{\delta T \rightarrow 0} \frac{\langle Q^r \rangle} {\delta T}.
	\end{equation} 
	For NESS systems the right hand side of Eq.~\ref{eq:C} is (negative) infinite because of the positive heat dissipation rate. 
	Boksenbojm {\it et al.}~\cite{Boksenbojm2011} could obtain a finite heat capacity by subtracting the housekeeping heat from the total heat and using the resulting excess heat in Eq.~\ref{eq:C}. 
	The present paper follows the same strategy; however, instead of following the Oono and Paniconi (OP) scheme to define the housekeeping and excess heat, we are going to follow the Hatano and Sasa (HS) scheme~\cite{Hatano2001, Ge2010}.

	The total heat $\langle Q^r \rangle$ is given by the integral
	\begin{equation}
	\label{eq:Total}
	\langle Q^r \rangle =  \int_{0}^{\infty} \!\!\! \mathrm{d}t \, \sum_i p_i(t) \sum_{j \neq i} R_{ji} \, ( E_i - E_j + T \Delta S_{ji}^{r, ext}). 
	\end{equation}
	The rationale of the integrand (instantaneous heat dissipation rate at time $t$) is the following: at any time $t$, the system is in configuration $i$ with probability $p_i(t)$; from $i$, the rate of transition to $j$ is $R_{ji}$; and for each such transition an amount of heat $(E_i - E_j) + T \Delta S_{ji}^{ext}$ is transferred to the reservoir. 
	To define the housekeeping heat, it is fruitful to consider the notion of effective {\it nonequilibrium forces}, $\{F_{ij}^{neq}(T)\}$, defined by the following relations~\cite{Esposito2010, Maes2013}
	\begin{equation}
	\label{eq:Force}
	F^{neq}_{ij}(T) = \frac{T}{2} \ln{\left[ \frac{R_{ij} \, p_j^S(T)}{R_{ji} \, p_i^S(T)}\right]}.
	\end{equation}  
	The term ``nonequilibrium" is used to denote the fact that NESS systems can be characterized by the presence of at least one nonzero $F_{ij}^{neq}(T)$.
	The average housekeeping heat in the HS approach, $\langle Q_{hk}^{r, HS} \rangle$, is the average work done by these nonequilibrium forces over the relaxation period~\cite{Ge2010},
	\begin{equation}
	\label{eq:Qhkdot}
	\langle Q^{r, HS}_{hk} \rangle = \int_0^{\infty} \!\!\! \mathrm{d}t \, \sum_{i, \, j \neq i} F^{neq}_{ij}(T) \, J_{ij}(t).
	\end{equation}
	This force-flux formula for housekeeping heat is reminiscent of a similar formula for entropy production rate in linear irreversible thermodynamics~\cite{deGroot1962}. 	
	Note also that the same formula gives the housekeepog heat in the OP approach of Boksenbojm {\it et al.}~\cite{Boksenbojm2011} if the instantaneous fluxes $J_{ij}(t)$ are replaces by final steady state fluxes $J_{ij}^S$.
	The average excess heat in the HS approach, $\langle Q_{ex}^{r, HS}\rangle$, is then given by the difference
	\begin{equation}
	\label{eq:Heats}
	\langle Q_{ex}^{r, HS} \rangle  =  \langle Q^r\rangle - \langle Q^{r, HS}_{hk}\rangle.
	\end{equation}
	Correspondingly, NESS heat capacity is defined as
	\begin{equation}
	\label{eq:HeatCap}
	C_{ex}^{HS}(T) = - \lim_{\delta T \rightarrow 0} \frac{\langle Q^{r, HS}_{ex} \rangle}{\delta T}.
	\end{equation} 
	In the following we shall see that the NESS heat capacity defined in Eq.~\ref{eq:HeatCap} is finite, and is proportional to the temperature derivative of the steady state Shannon entropy, 
	\begin{equation}
	\label{eq:Shannon}
	S^S(T) = - \sum_i p_i^S(T) \ln{p_i^S(T)}.
	\end{equation}
	
	{\it Derivation. -- }
	We can combine Eqs.~\ref{eq:Flux}, \ref{eq:Total}, \ref{eq:Force}, \ref{eq:Qhkdot} and \ref{eq:Heats} to get the following expression for excess heat~\cite{Ge2009, Ge2010, Seifert2012}
	\begin{equation}
	\label{eq:Int1}
	\langle Q^{r, HS}_{ex} \rangle  =  - T \int \! \mathrm{d}t \, \sum_i p_i(t) \sum_{j \neq i} R_{ji} \ln{\left[ \frac{p_i^S(T)}{p_j^S(T)}\right]}.
	\end{equation}
	This can be simplified further:
	\begin{eqnarray}
	\label{eq:Int2}
	\langle Q^{r, HS}_{ex} \rangle  & = &   T  \int  \mathrm{d}t \, \sum_i \ln{p_i^S(T)} \sum_{j \neq i} J_{ij}(t) \nonumber \\
	& = &  T  \int  \mathrm{d}t \, \sum_i \ln{p_i^S(T)} \left[ \frac{\mathrm{d}}{\mathrm{d}t} p_i(t) \right]  \nonumber \\
	 & = &  T  \sum_i  \ln{p_i^S(T)} \, \delta p_i^S(T)
	\end{eqnarray}
with $\delta p_i^S (T) = p_i^S(T) - p_i^S(T - \delta T).$
	 The first line is obtained from Eq.~\ref{eq:Int1} by a rearrangement of terms; the second line follows by using the master equation~\ref{eq:MasterCurrent}; and the last line is obtained by performing the integration over time.  
	Using Taylor expansion for $\delta p_i^S (T)$ up to linear order in $\delta T$, i.e., $ \delta p_i^S (T) \approx \delta T \frac{ \partial}{\partial T} \, p_i^S(T)$, we can further get
	\begin{eqnarray}
	\label{eq:Shannon_Simple}
	\langle Q^{r, HS}_{ex} \rangle & \approx & T \, \delta T \sum_i \ln{p_i^S(T)}\, \frac{\partial}{\partial T}\,  p_i^S(T)\nonumber \\
	& = & T \, \delta T \frac{\partial}{\partial T} \sum_i p_i^S(T) \ln{p_i^S(T)}  \nonumber \\
	& = & - \delta T \left[ T \frac{\partial}{\partial T} S^S(T) \right]. 
	\end{eqnarray}
	In the second line we have used the normalization condition, $\sum_i p_i^S(T) = 1$, to bring the derivative with respect to temperature outside of the summation, and in the third line we have used the definition of Shannon entropy, Eq.~\ref{eq:Shannon}. 
	Combining Eqs.~\ref{eq:HeatCap} and \ref{eq:Shannon_Simple} we get the final result
	\begin{equation}
	\label{eq:main}
	C_{ex}^{HS}(T) = T \frac{\partial}{\partial T} S^S(T).
	\end{equation} 
	Exactly the same relation holds between equilibrium heat capacity and entropy~\cite{Callen1985}. 
	In comparison, heat capacity in the OP approach leads to~\cite{Boksenbojm2011, Sagawa2011}
	\begin{equation}
	\label{eq:CexOP}
	C_{ex}^{OP}(T) = \frac{\partial}{\partial T} \langle E \rangle^S - \sum_{ijk} \Delta S_{ij}^{r, ext} R_{ij} R^\dagger_{jk} \frac{\partial}{\partial T} p_k^S(T),
	\end{equation}
where $\langle E \rangle^S$ is the average steady state energy and $R^\dagger$ is the Moore-Penrose pseudo inverse~\cite{Campbell1991} of the matrix formed by the rates $\{R_{ij}\}$~\cite{vanKampen2007}.

	(In an alternate strategy, authors in Ref.~\cite{Zia2002} considered only the first term on the right of Eq.~\ref{eq:CexOP} as the definition of NESS heat capacity. 
	It will be interesting to see if this approach is equivalent to any calorimetric definition.)

	{\bf One dimensional diffusion with periodic boundary condition. -- } 
	Consider now a particle constrained to move on a circle and in contact with a thermal reservoir at absolute temperature $T$.
	Let its angular position be denoted by $\theta$.
	If the particle is subject to a periodic potential $V(\theta) $ and a constant external torque $f_{ext}$ its Langevin equation of motion, in the over-damped limit, can be written as~\cite{vanKampen2007}
	\begin{equation}
	\label{eq:Langevin}
	\gamma \, \dot{\theta} = - V'(\theta) + f_{ext} + \eta (t).
	\end{equation}
	Here, $\gamma$ denotes the dissipation coefficient; the dot over $\theta$ denotes its time-derivative; the prime over potential $V(\theta)$ denotes its angular derivative; and $\eta(t)$ denotes Gaussian white noise with zero mean and delta function auto-correlation, $\langle \eta(t) \eta(t') \rangle = 2 \gamma  T \delta(t - t') $.  
 	(We have assumed both the mass of the particle and the radius of the circle to be unity.)
	The time evolution of the probability density function $\rho(\theta, t)$ is given by the Fokker--Planck equation~\cite{Risken1984, vanKampen2007}
	\begin{equation}
	\label{eq:FokkerPlanck}
	\dot{\rho}(\theta, t) = - J'(\theta, t) \quad, \quad J(\theta, t) = \frac{1}{\gamma} (f_{ext} - V') \rho - \frac{ T}{\gamma}\rho', 
	\end{equation}
where the function $J(\theta, t)$ is the probability flux at position $\theta$ at time $t$. 
	Because of the periodic nature of the system, density $\rho(\theta, t)$ and its derivatives must also satisfy periodic boundary conditions; in particular, we must have $\rho(\theta + 2 \pi, t) = \rho(\theta, t)$ and $J(\theta + 2 \pi, t) =  J(\theta, t)$.

	Consider now the same process as before: initiating the system in its steady state at temperature $(T - \delta T)$, and then suddenly bringing it in contact with a reservoir of temperature $T$ and letting it relax. 
	The average heat transferred to the environment, $\langle Q^r\rangle$, is the negative of the average work done by the environmental torque on the system~\cite{Sekimoto1998}, $( - \gamma \dot{\theta} + \eta ),$ i.e.,
	\begin{equation}
	\label{eq:Sekimoto}
	\langle Q^r\rangle  =  - \left\langle \int_{0}^{\infty} \!\! \mathrm{d}t \,  \left[ - \gamma \dot{\theta}(t) + \eta(t) \right] \dot{\theta}(t) \right\rangle.
	\end{equation}
	The time integral above has to be interpreted in the Stratonovich sense~\cite{Seifert2012}.
	Using Eq.~\ref{eq:Langevin} and the fact that the conditional average velocity at position $\theta$ at time $t$, $\langle \dot{\theta}(t) | \theta, t \rangle$,  is given by the following relation (again, in the Stratonovich sense)~\cite{Seifert2012} 
	\begin{equation}
	\label{eq:Local_Velocity}
	\langle \dot{\theta}(t) | \theta, t \rangle = J(\theta, t) / \rho(\theta, t),
	\end{equation}
we can rewrite Eq.~\ref{eq:Sekimoto} as~\cite{Ge2009}
	\begin{equation}
	\label{eq:Qtot_Langevin}
	\langle Q^r\rangle = \int_{0}^{\infty} \! \! \! \mathrm{d}t \, \int_{0}^{2 \pi} \! \! \mathrm{d}\theta \, \left[ f_{ext} - V'(\theta) \right] J(\theta, t).
	\end{equation}
	The nonequilibrium force at any position $\theta$ is~\cite{VandenBroeck2010, Maes2012}  
	\begin{equation}
	\label{eq:Force_Langevin}
	F^{neq}(\theta; T) = f_{ext} - V'(\theta) - T \left[ \ln{\rho^S(\theta; T)}\right]'.
	\end{equation}
	Note that $F^{neq}(\theta)$ is zero whenever the nonconservative force $f_{ext}$ is zero, but they are not proportional to each other. 
	In fact, $F^{neq}(\theta)$ depends on the complete dynamics of the system via the term $- T [\ln{\rho^S(\theta; T)}]'$, reflecting the nonlocal nature of NESS~\cite{Landauer1975}.
	As before, the average housekeeping heat $\langle Q_{hk}^{r, HS} \rangle$ is given by the average work done by the nonequilibrium force,
	\begin{equation}
	\label{eq:Qhk_Langevin}
	\langle Q_{hk}^{r, HS}\rangle = \int_0^\infty \!\!\! \mathrm{d}t \int_0^{2 \pi} \!\!\! \mathrm{d}\theta \, F^{neq}(\theta; T) \, J(\theta, t). 
	\end{equation}
	The average excess heat $\langle Q_{ex}^{r, HS}\rangle$ is then obtained from Eq.~\ref{eq:Heats}.
	We will now consider a brief derivation of Eq.~\ref{eq:main} for the present case. 
	The limits of integration, and sometimes the argument $\theta$, will be omitted to avoid clutter.

	{\it Derivation. -- }
	Combining Eqs.~\ref{eq:Heats}, \ref{eq:Qtot_Langevin}, \ref{eq:Force_Langevin} and \ref{eq:Qhk_Langevin} we get
	\begin{equation}
	\label{eq:Qex_Langevin}
	\langle Q_{ex}^{r, HS}\rangle  =   T  \int \mathrm{d}t \int \mathrm{d}\theta \, \left[ \ln{\rho^S(\theta; T)}\right]' J(\theta, t).
	\end{equation}
This implies that the excess heat $\langle Q_{ex}^{r, HS} \rangle$ is the average work done by the gradient force derived from the steady state distribution $\rho^S(\theta; T)$. 
	We can simplify Eq.~\ref{eq:Qex_Langevin} further: 
	\begin{eqnarray}
	\label{eq:Qex_Langevin_Simpl}
	\langle Q_{ex}^{r, HS}\rangle  & = &  T  \int \mathrm{d}t \int \mathrm{d}\theta \, \ln{\rho^S(T)} \, [ - J'(t)] \nonumber \\
	& = &  T  \int \mathrm{d}t \int \mathrm{d}\theta \,  \ln{\rho^S(T)} \, \dot{\rho}(t) \nonumber \\
	& = &   T \! \int \! \mathrm{d}\theta \, \ln{\rho^S(T)} \, \delta \rho^S(T)
	\end{eqnarray}
with $ \delta \rho^S(T) = \rho^S(T) - \rho^S(T - \delta T).$
	In the first line, we have integrated by parts over $\theta$ and used the periodic boundary conditions on $\rho(\theta, t)$ and $J(\theta, t)$; in the second line, we have used the Fokker-Planck equation~\ref{eq:FokkerPlanck}; and in the last line, we have performed the time integration on $\dot{\rho}(t)$. 
	The rest of the derivation is just the continuous version of the steps in Eq.~\ref{eq:Shannon_Simple} and is omitted here.

	By a slight generalization of the above derivations it can be shown that the following equality holds for an arbitrary quasistatic process (involving variations in any parameter, not just the temperature)~\cite{Hatano2001, Ge2009, Maes2013}
	\begin{equation}
	\int \frac{\langle Q_{ex}^{r, HS} \rangle}{T} = - \Delta S^S \quad, \quad \Delta S^S =  S^S(\text{final}) - S^S(\text{initial}).
	\end{equation}
	Thus, the inverse of absolute temperature $T$ acts as the integrating factor for Hatano Sasa excess heat $\langle Q_{ex}^{r, HS} \rangle$ for an arbitrary quasistatic process and leads to the state function Shannon entropy.
	This is an exact, nonequilibrium generalization of the Clausius approach to equilibrium entropy~\cite{Hatano2001, Ge2009, Maes2013}, because, in equilibrium, Hatano Sasa excess heat $\langle Q_{ex}^{r, HS} \rangle$ is the total heat and Shannon entropy is the thermodynamic entropy.
	Furthermore, the general Clausius inequality holds for an arbitrary process~\cite{Hatano2001, Ge2009, Maes2013}:
	\begin{equation}
	\int \frac{\langle Q_{ex}^{r, HS} \rangle}{T} \geq - \Delta S^S.
	\end{equation}	
	This strongly implies the perspective that the Shannon entropy is the natural NESS generalization of equilibrium entropy and, therefore, the heat capacity in Eqs.~\ref{eq:HeatCap} and \ref{eq:main} is the natural NESS generalization of equilibrium heat capacity.  
	In particular, the relation~\ref{eq:main} is process-independent, just like its equilibrium counterpart.

	   I acknowledge many useful discussions with Christopher Jarzynski and Sebastian Deffner, and financial support from the National Science Foundation (USA) under grants DMR 1206971 and OCE 1245944.

\end{document}